\documentstyle[12pt,aaspp]{article}

\lefthead{PAGE, SHIBANOV \& ZAVLIN}
\righthead{GEMINGA: STRUCTURE OF THE SURFACE}

\newcommand{\be}{\begin{equation}}
\newcommand{\ee}{\end{equation}}
\newcommand{\Msol}{\mbox{$M_{\odot}\;$}}
\def\lsim{\lower 2pt \hbox{$\, \buildrel {\scriptstyle <}\over
         {\scriptstyle \sim}\,$}}

\begin{document}

\title{GEMINGA'S SOFT X-RAY EMISSION AND \\
       THE STRUCTURE OF ITS SURFACE}
\author{Dany Page}
\affil{Instituto de Astronom\'{\i}a, UNAM,
       Apdo Postal 70-264,
       04510 M\'{e}xico D.F., M\'{e}xico. \\
       page@astroscu.unam.mx}
\author{Yu.~A. Shibanov}
\affil{Ioffe Physical-Technical Institute,
       St Petersburg 194021, Russia. \\
       shib@ammp.ioffe.rssi.su}
\and
\author{V.~E. Zavlin}
\affil{Max Planck Institut f\"{u}r Extraterrestrische Physik,
       Garching, Germany. \\
       zavlin@rosat.mpe-garching.mpg.de}

\begin{abstract}

  We present a model to explain the decrease in the amplitude of the
  pulse profile with increasing energy observed in Geminga's soft
  X-ray surface thermal emission.  We assume the presence of plates
  surrounded by a surface with very distinct physical properties:
  these two regions emit spectra of very distinct shapes which present
  a crossover, the warm plates emitting a softer spectrum than the
  colder surrounding surface.  The strongly pulsed emission from the
  plates dominates at low energy while the surroundings emission
  dominates at high energy, producing naturally a strong decrease in
  the pulsed fraction.  In our illustrative example the plates are
  assumed to be magnetized while the rest of the surface is field
  free.

  This plate structure may be seen as a schematic representation of
  a continuous but very nonuniform distribution of the surface magnetic
  field or as a quasi realistic structure induced by past tectonic
  activity on Geminga.

\end{abstract}


\bigskip
\bigskip
Submitted to {\em The Astrophysical Journal Letter.}

\newpage
\section{INTRODUCTION}

Despite of the large amount of observational data available on radio
pulsars, and more recently on X-ray and $\gamma$-ray pulsars, the
structure of neutron stars still remains ellusive.  Not only the
interior is a puzzle, but even the structure of the very surface is a
mystery: in the low density - low temperature region near the surface
the effects of the magnetic field on the structure of matter and the
equation of state still defy any conclusive description.  Early work
on this question claimed that an iron surface should be a magnetic
solid (Ruderman 1974) instead of an atmosphere.  Improved calculations
in this direction showed that the problem is extremely delicate and
the situation is not yet clearly resolved (see, e.~g., Neuhauser {\em
  et al.} 1986 and K\"{o}ssl {\em et al.} 1988).  However, the surface
itself, or the atmosphere, is not necessarily made of iron but
probably consists of light elements. A few grams per $\rm cm^2$ of
hydrogen at the surface is sufficient to give an optical depth of
unity and could be present due to accretion from the interstellar
medium or produced on site from the original iron layer by the
bombarding of highly energetic particles generated in the
magnetosphere.  Moreover, fallback of matter after the supernova
explosion can also cover the surface, and its magnetic field, by a
layer of light elements (Chevalier 1989; Muslimov \& Page 1995).  If
hydrogen is present, the heavier elements will sediment due to the
enormous gravitational field: this would reduce the study of the
surface to the comparatively simpler case of a magnetized hydrogen
atmosphere.  However, hydrogen may agregate into molecules (Lai {\em
  et al.} 1992) because of the magnetic field, and convection within
the atmosphere at low temperature ($T \lsim 10^5$ K, Zavlin {\em et
  al.} 1995a) may mix the hydrogen with the heavier elements laying
beneath it, complicating again the problem.

Thank to deep {\em ROSAT} observations, we now have strong evidence
that thermal radiation from four nearby neutron stars has been
detected (\"{O}gelman 1995): we are at long last {\em seing} the
surface of a neutron star.  Fitting of these observed thermal spectra
with theoretical spectra rised the hope of allowing us to select the
correct atmosphere or surface model(s) for these neutron stars.
Unfortunately, the energy resolution of the {\em ROSAT} PSPC detector
and unsolved problems in the calibration of this detector at low
energy prevented this hope to materialize since analyses with various
spectra have given equivalent results: {\em spectral fits by
themselves are not sufficient to clearly discriminate among the
various atmosphere models} (Page 1995b; Meyer {\em et al.} 1994)
with the present observational capability.  On the other hand, the
strong modulation of the observed pulsed profiles can be interpreted
as due magnetic field effects inducing large surface temperature
differences (Page 1995a, 1995b, Page \& Sarmiento 1995) and/or
anisotropy of the atmospheric emission (Shibanov {\em et al.} 1994;
Zavlin {\em et al.} 1995b).  These works thus showed that the shape of
the light curves and their energy dependence contain crucial
information.

Presently, the most intriguing and unexplained observed feature is
found in Geminga (Halpern \& Ruderman 1993): the amplitude of the
pulsations (or pulsed fraction, $Pf$) in the PSPC channels 8 to 28
(i.e., roughly at energies below 300 eV) is much larger than in
channels 28 to 53: 33\% vs. 20\%.  Blackbody emission {\em always}
gives an increase of $Pf$ with energy (Page 1995a, Page \& Sarmiento
1995). This `Geminga effect' thus requires the inclusion of magnetic
effect on the emitted spectrum and not only on the surface temperature
distribution.  The preliminary results of Shibanov {\em et al.} (1994)
showed that realistic atmospheric models are able to produce a slight
decrease of $Pf$ with increasing energy, but still much smaller than
what is observed.

We present here a simple model which is able to produce this decrease
in $Pf$ in a very natural way by assuming that the surface of Geminga
consists of regions with very distinct physical properties (\S 2).
Our results are presented in \S 3. We discuss in \S 4 some possible
reasons for such a surface structure and conclude in \S 5.

\section{THE MODEL}

We model the surface of the Geminga pulsar as made of two uniformly
magnetized plates, containing the `north' and `south' magnetic poles,
surrounded by a nonmagnetized crust.  This can be considered either as
a quasi realistic structure within the scenario of plate tectonics
(see \S 4) or as a schematic representation of a continuous but very
nonuniform distribution of surface magnetic field.  It has the
advantage of simplicity, requiring only two models of atmosphere.
Considering Geminga as an orthogonal rotator (Halpern \& Ruderman
1993), the two magnetic plates are located near to the rotational
equator.  If the surface temperature is determined by the heat flow
from the hot interior through the crust, the magnetized plates must be
warmer than the remainder of the surface (Page 1995a) and the presence
of a single peak in the observed soft X-ray light curves imply that
the plates are close to each other.  We assume that the whole stellar
surface is covered by a hydrogen atmosphere, magnetized on the plates.
The plates and the off-plate region have thus very distinct emission
properties: the off-plate region has a much harder spectrum, i.e., a
large excess in the Wien tail, compared to the plate spectrum due to
the different frequency dependence of the opacity (Shibanov {\em et
  al.} 1992).  We use the emission spectra calculated by Pavlov {\em
  et al.} (1994) which are similar to the spectra used to perform
spectral fit for Geminga by Meyer {\em et al.} (1994).

The parameters of our model are thus the followings:
\begin{itemize}
\item \vspace{-16pt}
Diameter and position of the two magnetic plates.
\item \vspace{-8pt}
Temperature of the off-plate region $T_{op}$ and magnetic field $B_p$
and temperature $T_p$ of the plates.
\item \vspace{-8pt}
Mass $M$ and radius $R$ of the star.
\item \vspace{-8pt}
Distance $D$ of the star, interstellar column density $N_H$ and
orientation of the observer with respect to the star's rotation axis
$\zeta$.
\end{itemize}
 \vspace{-16pt}
We fix $M$ at 1.4 \Msol and $R$ at 10 km and take $\zeta = 90^\circ$.
The temperature $T_{op}$ (and to some extent  $T_p$) as well as $D$ and $N_H$
are determined by fitting the spectrum while the size and location of the
plates and $B_p$ and $T_p$ are determined mainly by fitting the shape of
the light curves.

\section{RESULTS}

The values of the parameters are found in Fig.~1 which
shows the resulting spectrum: the dominance of the plate emission at
low energy is clearly seen.  Since the plate emission is the cause of
the pulsations this naturally implies that the pulsed fraction is
strongly decreasing with increasing photon energy as can be seen in
Fig.~2.  This result depends critically on the presence of
(at least) two different emitting regions with a {\em crossover} in
their spectra.  This crossover is due here to the effect of the
magnetic field on the plate spectrum which is absent off-plate but
other mechanims providing the same spectral properties would give the
same results. The most interesting point is that the warm region (the
plates) must have a softer spectrum than the cold one.  This is the
contrary of what one would expect if the whole surface of the star
were homogenously magnetized or non magnetized: the cold atmosphere
would show more intense absorption and have a softer spectrum than the
warm region.  Our result requires thus a drastic difference in the
physical properties of the warm region compared to the cold one, here
the presence or absence of the magnetic field.

We show finally in Fig.~3 our fit to the light curves in
three channel ranges: the decrease of the amplitude of the pulsations
with increasing channel number is clearly seen again in bands 7-28 vs.
28-53.  The higher channel band 53-256 (not shown in the figure)
corresponds to the hard tail which has a different origin than the
thermal emission from the whole surface (Halpern \& Ruderman 1993) and
we do not attempt to model it.  It was proposed by Halpern \& Ruderman
(1993) that the decrease in the pulsed fraction of the surface thermal
component may be caused by contamination from this hard tail, whose
pulses are $105^\circ$ off-phase with respect to the thermal
component.  However, the contribution of the hard tail in the channel
range 28-53 is one to two orders of magnitude below the contribution
from the thermal component, depending if the hard tail is modeled as a
power law or blackbody spectrum (see Fig. 3 and 6 in Halpern \& Ruderman
1993) and cannot possibly be responsible for the decrease in the
pulsed fraction (Page \& Sarmiento 1995).

The above results also give us a clue to the behaviour of the pulsed
fraction in other models. The increase in $Pf$ obtained with
blackbody spectra (Page 1995a, Page \& Sarmiento 1995) is most certainly
due to the fact that the blackbody hardness increases with
temperature. The small decrease in $Pf$ obtained with magnetized
hydrogen spectra and a dipolar surface field (Shibanov {\em et al.}
1994) is itself due to the slight softening of these spectra with
increasing field strentgh and the fact that the regions with stronger
field have a higher temperature.

\section{DISCUSSION}

\subsection{The parameters of the model}

We have obtained similar results using a larger field strength ($\sim
10^{13}$ G) or even by using a blackbody spectrum for the plates: both
give the same crossover with the non magnetic spectrum at roughly the
same energy. The distance $D$ is then larger ($\sim$ 30 pc), with the
same $N_H$.  The distance $D$ and $N_H$ we obtain are comparable to
the ones obtained by Meyer {\em et al.} (1994) to whom we refer for a
discussion of this aspect of the problem.  The exact values of the
parameters moreover depend on the detector's response. We have used
the 1992 March 19 response matrix and the 1993 January 12 version implies,
e.g., changes of the order of 5\% in temperature and and increase of
about 15\% in $N_H$.
We did not attempt to perform an accurate spectral fit but only tried
to show the main possible mechanism able to produce the `Geminga
effect' as observed in the first set (March 1991) of {\em ROSAT}
observations.  Our spectral fit does have an excess in the channel
range 50-90: spectra softer than the non magnetized hydrogen one are
needed for a more complete study.  A detailed fit of {\em both} the
spectrum and the energy dependent light curves impose very strong
constraints on the atmosphere models and models presently available
are apparently not sufficient.

The ratio of the onplate to offplate temperatures we need is within
the predicted range from models of heat transport in magnetized and
non magnetized neutron star envelopes (Van Riper 1988). However at
such low surface temperatures the envelope models are extremely
uncertain due to the dominating effect of the magnetic field on the
equation of state; magnetized atmosphere models are also not very
reliable, since the effect of the atomic motion in the magnetic field
have not yet been included (Pavlov \& M\'{e}sz\'{a}ros 1993).

\newpage
\subsection{Reasons for surface inhomogeneity}

The large inhomogenity needed in our model could be attributed to
chemical changes at the surface if the surface is solid. If the
surface is gaseous or liquid (covered or not by an atmosphere) then
meridional flows, induced by the rotation and/or the magnetic field
gradient, would very probably homogenize the chemical composition;
in this case only magnetic field inhomogeneity could be invoked to
produce the needed spectra.

Magnetic field inhomogeneity is thus most probably the major agent in
inducing large variation in the emitted spectrum.  In our example we
invoked regions with and without magnetic field.  Other possibilities
are the formation of a magnetic solid surface or the presence of
magnetic molecules, both depending on the local field strength and
temperature (itself depending on the strength of the underlying field).  The
strong magnetic inhomogeneity may be due, e.g., to the upraising of
the interior field by ohmic diffusion (Muslimov \& Page 1995) or to
early tectonic activity of the pulsar.

\subsection{Plate tectonics}

The spin-down of the pulsar induces tremendous stress on the crust
both by the differential rotation of the crust superfluid neutrons
and/or by the outward motion of the core neutron vortices.  Release of
this stress can occur either by braking of the crust or unpinning of
the crust neutron vortices.  The theory of plate tectonics (Ruderman
1976, 1991a, b, c) stipulates that fast spinning pulsars can be
expected to break their crust before the crust vortices can unpin.
Crust breaking and vortex unpinning are natural explanations for
pulsar glitches (Ruderman 1976, 1991c; Alpar 1977) and analyses of the
Crab (Link {\em et al.} 1992; Alpar {\em et al.} 1994) and Vela (Alpar
{\em et al.} 1993) pulsar glitches indicate that crust braking is
occuring in the former while unpinning is sufficient to explain the
latter case.  These analyses strongly support the general pattern of
crust braking in young and/or fast spinning pulsars and vortex
unpinning in the slower ones.

As a consequence, neutron stars born with high spin rate can be expected
to have undergone a phase of crust braking and thus should keep on their
surface the marks of the previous platelets motion (toward the rotational
equator) and formation of a new crust between the moving platelets.
The mechanism of `injection' invoked in explaining pulsar statistics
(Vivekanand \& Narayan 1981) stipulates that many pulsars are born
with large periods: these pulsars may not have gone through a crust breaking
phase.
Thus the `marks' left by the tectonic activity are not necessarily present
in all pulsars.
However, if present, these marks should affect the properties of the
pulsar surface and its thermal emission, i.e.,
{\em they should be visible}.
The model we have presented here finds a natural explanation within
this theory of plate tectonic and the `Geminga effect' may be such a
sign of past tectonic activity on Geminga.

\section{CONCLUSIONS}

We have shown that the energy dependence of the amplitude of the
observed pulse profile of Geminga's soft X-ray thermal emission can be
naturally explained by using a superposition of two very different
spectra: a soft component which dominates at energies below 300 eV
and a harder component which dominates at higher energies (at energies
above $\sim$ 500 eV the surface thermal emission is hidden behind the
hard tail of different origin). The soft component produces the
pulsations, i.e., it is emitted by a smaller area than the hard
component, and hence the pulsed fraction naturally decreases with
increasing photon energy.  Since the soft spectrum is emitted by a smaller
region than the hard spectrum it very probably implies a higher
temperature in this small area to produce the required crossover of these
two spectra.

This postulated surface structure may be simply seen as an idealized
representation of a smooth but highly nonuniform distribution of the
surface magnetic field or as a quasi realistic structure resulting from
a past era of tectonic activity on Geminga.

\acknowledgments 

We thank J.~H. Halpern for providing us with the {\em ROSAT} data of Geminga,
as well as G.~G. Pavlov, D.~G. Yakovlev and A. Sarmiento for discussions.
This work was supported by a DGAPA grant No. IN105794 at UNAM.
At the Ioffe Institute it was supported by
an ESO C\&EE grant No. A-01-068,
a RFFR grant No. 93-02-2916
as well as an INTAS grant No. 94-3834.
V.~E.~Z. is grateful to the Max Planck Institut f\"{u}r Extraterrestrische
Physik for financial support through the visitor program.
Yu.~A.~Sh. is indebted to UNAM for its hospitality.

\clearpage

\clearpage

\begin{figure}
\caption{
         Total spectrum and partial spectra from the two plates and
         the offplate region.
         The plate spectrum is much softer than the surrounding's spectrum
         and dominates at low energy: this results in a strong decrease
         in the pulsed fraction with increasing photon energy
         as shown in figure 2.
         The circular magnetized plates have diameters of $60^\circ$ and
         $85^\circ$ and their centers are located on the rotational
         equator at a distance of $140^\circ$ from each other.
         The magnetic field strength on both plates is $4.7 \cdot 10^{12}$ G.
         The temperature of the plates is $T_p = 4.17 \cdot 10^5$ K while
         the main surface is at $T_{op} = 2.18 \cdot 10^5$ K.
         The star is at a distance of 15 pc and the effective hydrogen
         column density is $N_H = 3 \cdot 10^{20} \; \rm cm^{-2}$.
         (The two absorption edges at 0.284 and 0.532 keV are due to
         interstellar C and O respectively).}
\end{figure}

\bigskip

\begin{figure}
\caption{
         Pulsed fraction of the model
         as detected through the {\em ROSAT} PSPC and
         as would be recorded by a perfect spectrometer.
         We used the 1992 March 19 version of the PSPC's response matrix.}
\end{figure}

\bigskip

\begin{figure}
\caption{
         Comparison of the theoretical light curves with the observed
         pulse profiles of Geminga (Halpern \& Ruderman 1993).
         Line styles are as in Figure 1.
         Notice the strong decrease in the pulsation amplitude
         in the channel range 28 - 53 compared to 7 - 28.}
\end{figure}

\end {document}
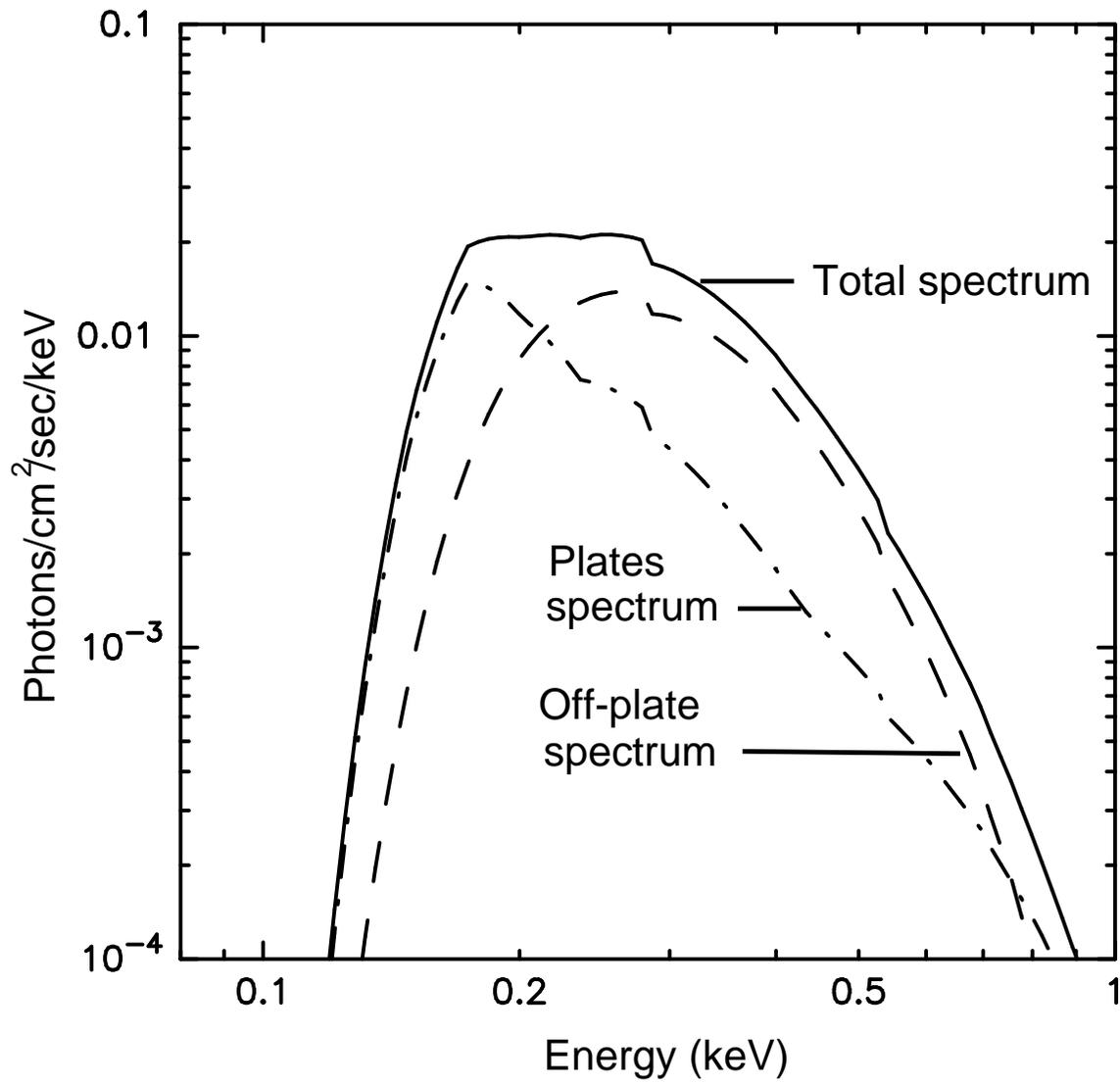
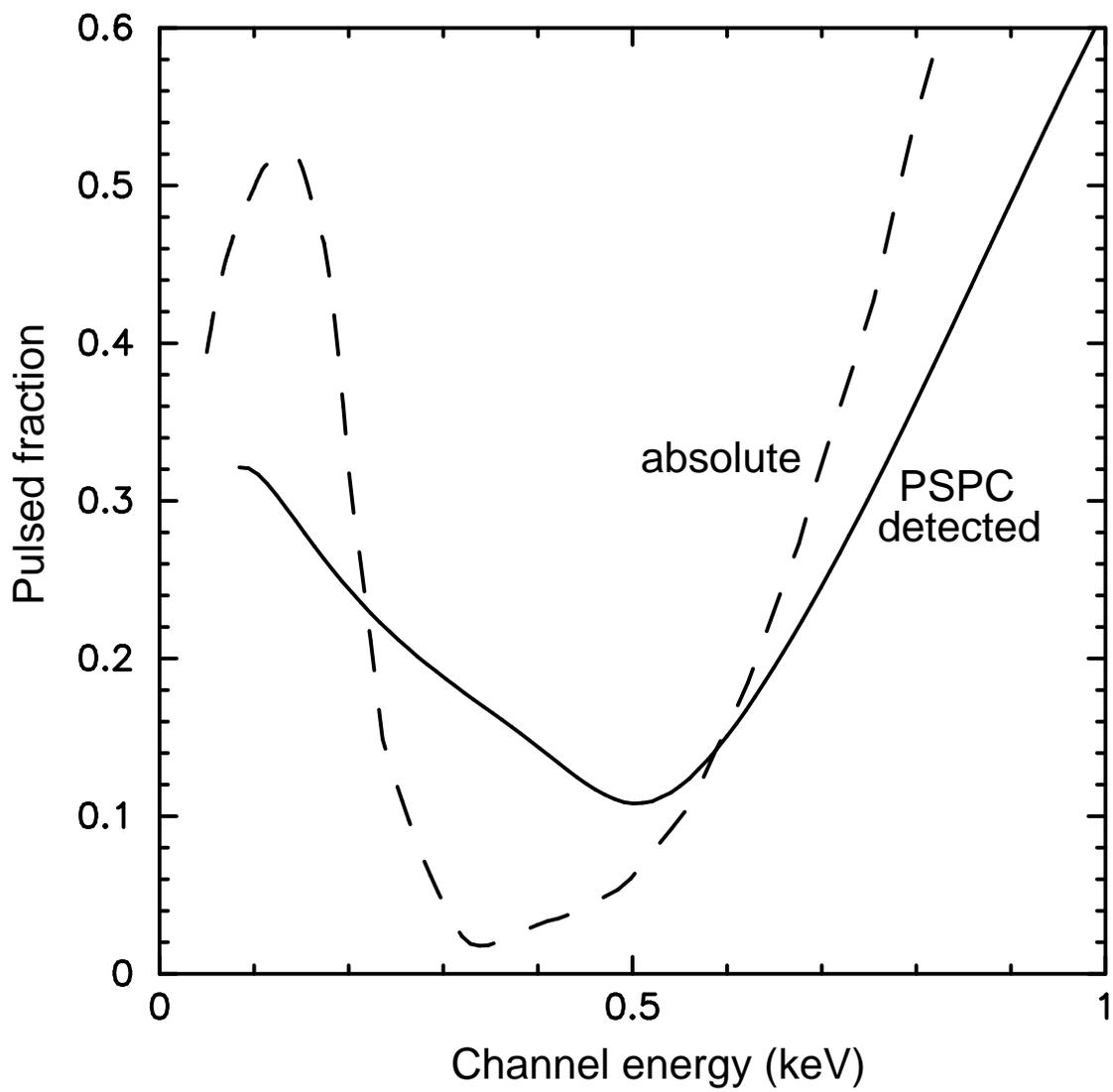
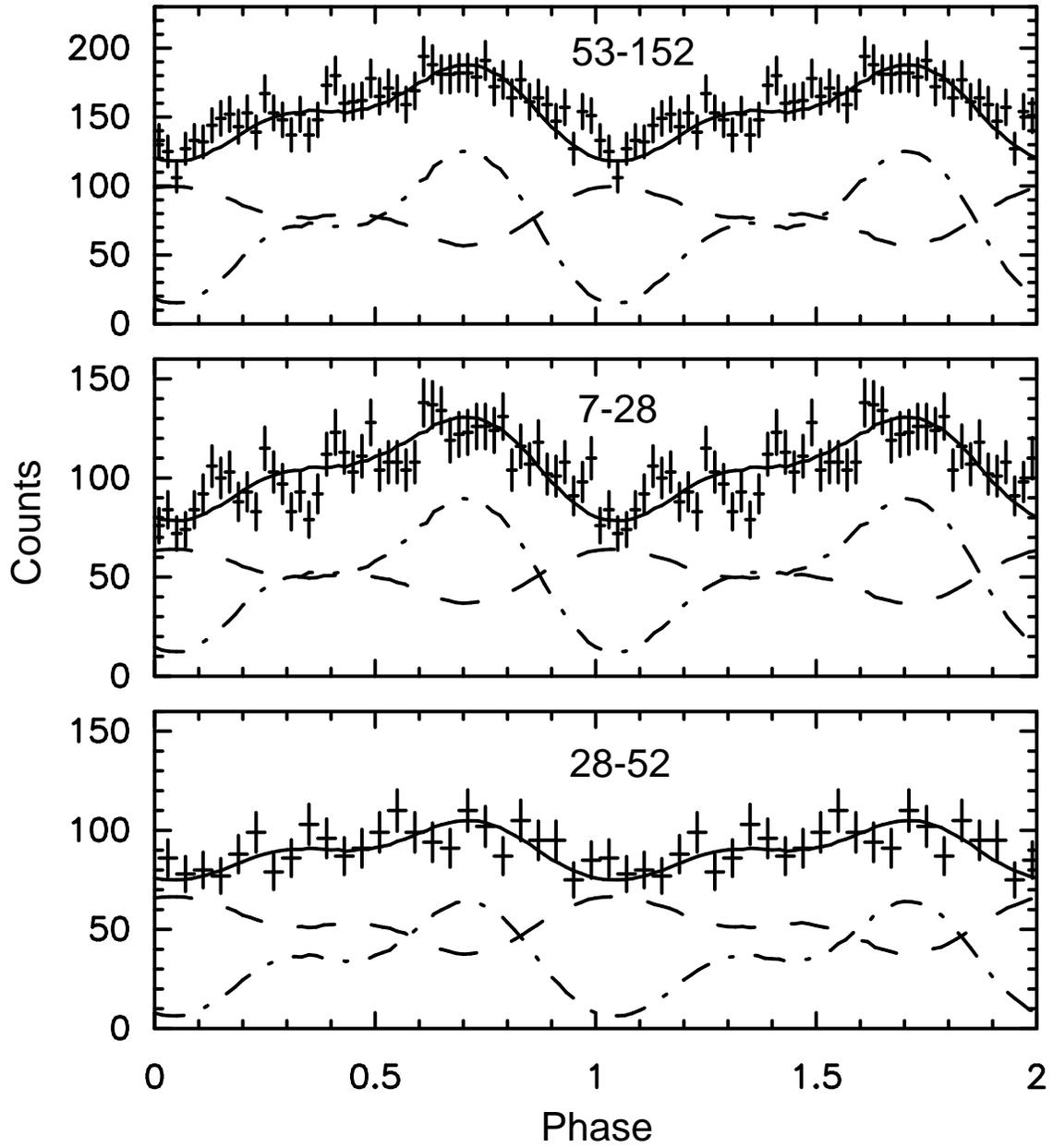
